\documentclass[aps,prl,twocolumn,amsmath,superscriptaddress]{revtex4-2}
\usepackage{CJK}
\usepackage{amssymb,amsmath}
\usepackage{graphicx,color}
\usepackage{lipsum}
\usepackage{ulem}
\usepackage{hyperref}
\usepackage{lineno}

\usepackage{ulem}

\begin{document}

\begin{CJK*}{UTF8}{gbsn}

\title{The yielding of granular matter is marginally stable and critical}

\author{Jin Shang}
    \affiliation{School of Physics and Astronomy, Shanghai Jiao Tong University, 800 Dong Chuan Road, 200240 Shanghai, China.}
\author{Yinqiao Wang}
    \affiliation{Research Center for Advanced Science and Technology, University of Tokyo, 4-6-1 Komaba, Meguro-ku, Tokyo 153-8505, Japan.}
\author{Yuliang Jin}
    \affiliation{Institute of Theoretical Physics, Chinese Academy of Sciences, Beijing 100190, China. }
    \affiliation{School of Physical Sciences, University of Chinese Academy of Sciences, Beijing 100049, China.}
    \affiliation{Wenzhou Institute, University of Chinese Academy of Sciences, Wenzhou 325000, China.}
\author{Jie Zhang}
    \email[Email address: ]{jiezhang2012@sjtu.edu.cn}
    \affiliation{School of Physics and Astronomy, Shanghai Jiao Tong University, 800 Dong Chuan Road, 200240 Shanghai, China.}
    \affiliation{Institute of Natural Sciences, Shanghai Jiao Tong University, 200240 Shanghai, China.}
    
\begin{abstract}
The mechanical yield of dense granular materials is a fascinating rheological phenomenon, 
beyond which stress no longer increases with strain at a sufficiently large deformation.
Understanding the behavior of mechanical responses associated with yielding is a fundamental goal in granular physics, and other related fields including glassy physics~\cite{Bonn-RevModPhys.89.035005,Liu-science.aai8830}, material sciences~\cite{greer2013shear}, geophysics~\cite{johnson2005nonlinear}, and active matter biophysics~\cite{manning-pnas-2021}.
However,  despite nearly half a century of theoretical efforts \cite{spaepen1977microscopic, argon1979plastic, Falk-Langer-PRE98, Falk-Langer-2011, Sollich-PRL-1997, Sollich-PRE-1998, Maloney-PRL-2004, Maloney-PRE-2006, Lematre-PRL-2009, Procaccia-PRE-2011, Procaccia-PRL-2012, Liu-2011, Liu-2015, Kawasaki2016PRE, Jaiswal2016PRL, Procaccia2017PRE, Parisi2017PNAS, Ozawa2018PNAS}, the nature of yielding in amorphous solids remains largely elusive compared to its crystalline counterpart. 
Here, we experimentally investigate the mechanical responses of two-dimensional bidisperse jammed disks subjected to volume-invariant pure shear, focusing on the behavior of yielding. 
We show that the microscopic mechanical and geometrical features of  configurations under shear can be characterized by two critical exponents of weak-force and small-gap distributions originally proposed for the isotropic jamming transition \cite{Charbonneau2014NC,Charbonneau2014JSMTE,Charbonneau2017ARCMP,Wyart2012PRL,Muller2015ARCMP}. 
We find that the yielding transition satisfies the condition of marginal mechanical stability through a scaling relationship between the two exponents, and after yielding global instability emerges.
The criticality of yielding is revealed by a significant peak of susceptibility that quantifies the fluctuation of a glass overlap order parameter.
Moreover, we find a distinct transition before yielding, which is associated with the onset of structural anisotropy.

\end{abstract}
\maketitle
\end{CJK*}

%%%%%%%%%%%%%%%%%%%%%%%%%%%%%%%%%%%%%%%%%%%%%%%%%%%%%%%%%%%%%%%%%%%%%%%%%%%%%%%%%%%%%%%
 \begin{figure*}
	\centerline{\includegraphics[width = 18 cm]{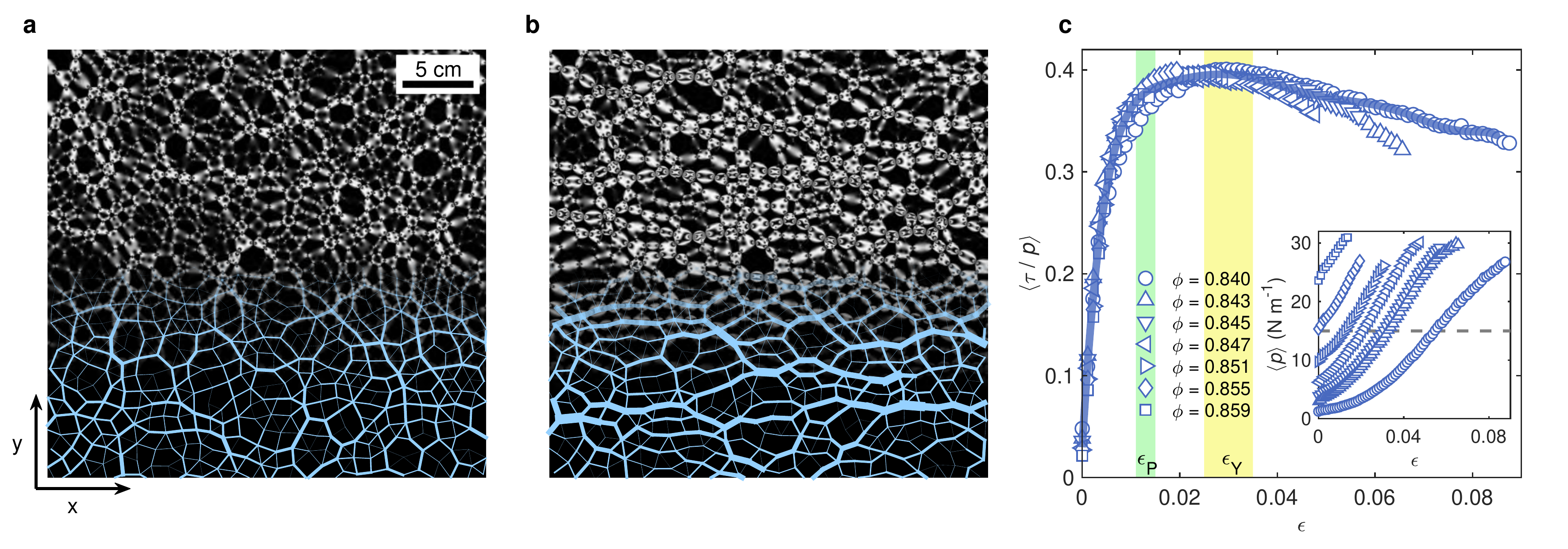}}
	\caption{
	Stacked stress and contact-force network images before (a) and after (b) pure shear (compression along the $x$ direction and expansion along the $y$ direction) under a constant packing fraction $\phi = 0.855$.  The upper halves show the raw photoelastic images. In contrast, the lower halves show the corresponding contact-force networks, whose edge's width is proportional to the contact-force magnitude. The shear strain $\epsilon$ in (b) equals $0.016$. (c) The global stress ratio $\left \langle \tau/p \right \rangle$ versus $\epsilon$ in systems of different $\phi's$. The data points of different $\phi's$ collapse nicely onto a single master curve, i.e., the solid line drawn for the guide to the eye. Inset: The average pressure $\left \langle p \right \rangle$ versus $\epsilon$ at different $\phi's$. The dashed line represents $p = 15 \ \rm{N \ m^{-1}}$, above which the critical exponents associated with weak forces and small-gaps can be measured accurately. The green and yellow vertical stripes represent approximately two characteristic points of $\epsilon_P \in [0.011,0.015]$ and $\epsilon_Y \in [0.025,0.035]$, respectively. 
	}
	\label{fig:figure1}
\end{figure*}

 \begin{figure*}
	\centerline{\includegraphics[width = 18 cm]{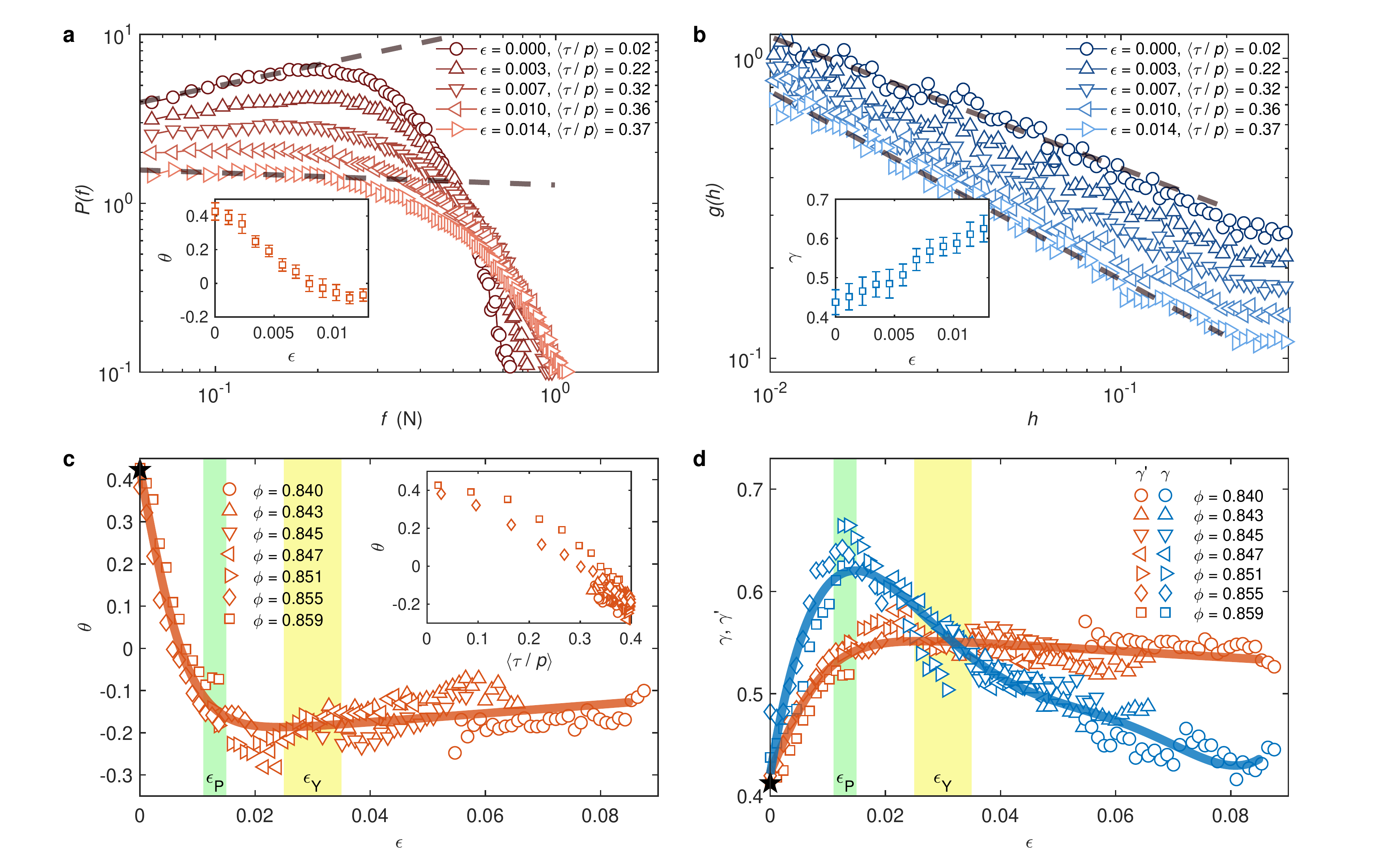}}
	\caption{
	(a) Contact-force distributions $P(f)$ of $\phi = 0.859$ at different shear strains $\epsilon's$. Inset: The weak-force critical exponents $\theta$ versus $\epsilon$.  (b) Gap distributions $g(h)$ of $\phi = 0.859$. Inset: The critical gap exponents $\gamma$ versus $\epsilon$. Two dashed lines in (a-b) represent the power-law fitting curves associated with the maximum and minimum strains, respectively. The error bars in the insets represent the standard deviations of the coefficients of the fitting parameters. (c) $\theta$ versus $\epsilon$ (main panel) and the stress ratio $\tau/p$ (inset) at different $\phi's$. (d) $\gamma$ and $\gamma'=1/(2+\theta)$ versus $\epsilon$ at different $\phi's$. The solid lines in (c-d) are a guide to the eye. The black stars refer to $\theta = 0.42311...$ and $\gamma = 0.41269...$ of the fullRSB theoretical prediction for an isotropic infinite dimensional hard-sphere system.
	}
	\label{fig:figure2}
\end{figure*}

\begin{figure}
	\centerline{\includegraphics[width = 8.6 cm]{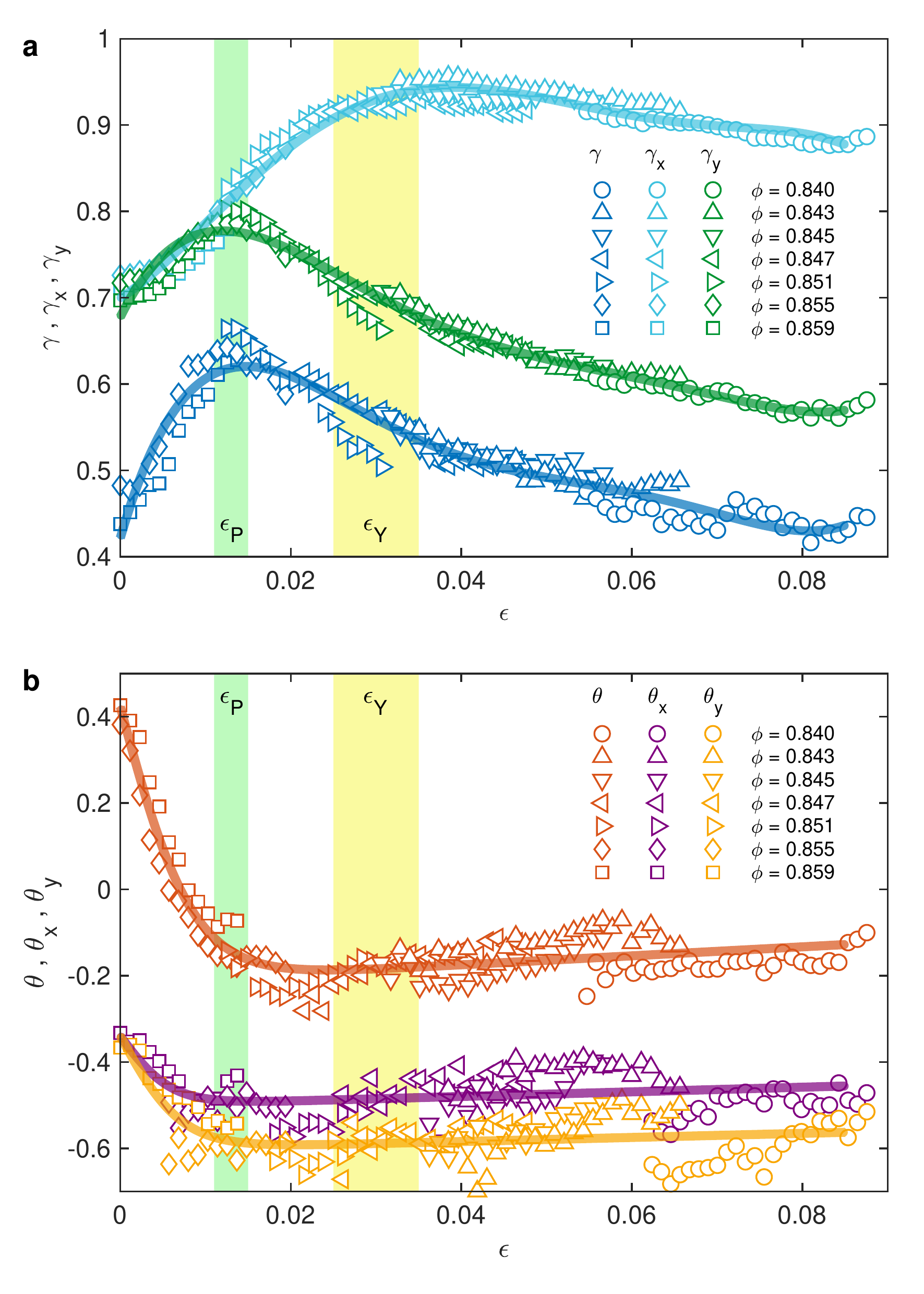}}
	\caption{
	(a) Critical gap exponents $\gamma$ and the associated $\gamma_x$ and $\gamma_y$ of the $x$ and $y$ components of gap versus $\epsilon$.  (b) Weak-force critical exponents $\theta$ and the associated $\theta_x$ and $\theta_y$ of the $x$ and $y$ components of weak forces versus $\epsilon$. Solid lines are a guide to the eye. 
	}
	\label{fig:figure3}
\end{figure}

 \begin{figure*} 
	\centerline{\includegraphics[width = 18 cm]{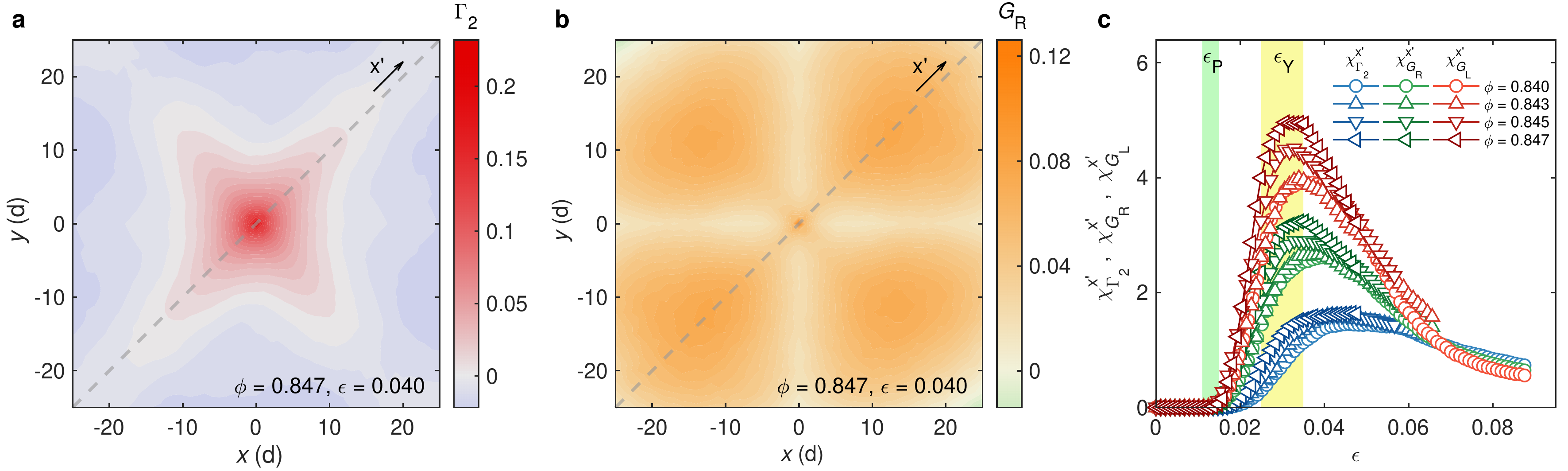}}
	\caption{
	Examples of the 2D correlation functions $\Gamma_2(x,y)$ (a) and $G_R(x,y)$ (b). The arrows mark the $x'$ direction, and the dashed lines represent $y'= 0$. The length unit $\mathrm{d}$ is the diameter of small disks. 
     (c) The susceptibilities $\chi^{x'}_{\Gamma_2}$, $\chi^{x'}_{G_R}$ and $\chi^{x'}_{G_L}$ as a function of the strain $\epsilon$ for different packing fractions $\phi's$. 
	}
	\label{fig:figure4}
\end{figure*}

%%%%%%%%%%%%%%%%%%%%%%%%%%%%%%%%%%%%%%%%%%%%%%%%%%%%%%%%%%%%%%%%%%%%%%%%%%%%%%%%%%%%%%% Main text
%%%%%%%%%%%%%%%%%%%% Introduction
\section*{I. Introduction}

When a dense granular material is subjected to small deformations, the responses are approximately elastic; if the strain further increases, many plastic events occur, and then the material undergoes a mechanical yield, with the appearance of system-spanning shear bands; after yielding, the system eventually enters into a steady-flow regime~\cite{sand-deformation,Zheng2018PRL}. These characteristics are universal in granular materials and other disordered solids, such as metallic and molecular glasses, emulsions, and colloidal glasses, prompting us to find a unified theoretical explanation. 
Most theories take a thermodynamic perspective: in early studies, the plastic deformation of amorphous solids is analyzed through shear-induced activation processes, under an effective temperature~\cite{spaepen1977microscopic,argon1979plastic, Falk-Langer-PRE98, Falk-Langer-2011, Sollich-PRL-1997, Sollich-PRE-1998} that evolves with strain; more recent developments propose that the mechanical yield is a thermodynamic phase transition~\cite{Rainone2015PRL, Parisi2017PNAS, Procaccia2017PRE, Jaiswal2016PRL, Kawasaki2016PRE}. 
Specifically,  mean-field theories predict yielding as a spinodal point in infinite dimensions \cite{Rainone2015PRL,Parisi2017PNAS,Procaccia2017PRE}, while simulations 
in physical dimensions suggest that a first-order-like transition occurs before the spinodal point \cite{Jaiswal2016PRL,Kawasaki2016PRE}. 
However, direct application and testing of the thermodynamic theories in granular experiments is a great challenge because inter-particle friction can often cause hysteresis and instability at the microscopic level.

% Mean-field and MMS
In a parallel and seemingly unrelated development, the marginal mechanical stability (MMS) analysis \cite{Wyart2012PRL,Muller2015ARCMP, lerner2013low, degiuli2014force} 
 focuses on the stability properties of isostatic random hard-sphere (or disk) packings that are jammed by isotropic compression.
 Such packings are characterized by power-law distributions of weak contact forces and small interparticle gaps, satisfying $P(f) \sim f^{\theta}$ and $g(h) \sim h^{-\gamma}$ respectively. The mechanical stability requires $\gamma \ge 1/(2+\theta)$ with the equality achieved for systems under the marginally stable condition.
The mean-field full replica symmetry breaking (fullRSB) glass theory predicts that $\theta = 0.42311...$ and $\gamma = 0.41269...$ in large dimensions~\cite{Charbonneau2014NC,Charbonneau2014JSMTE}, which have been verified in the recent experimental work \cite{Wang2022PNAS}. 
When anisotropy is introduced, it is reported that the two exponents remain the same as in the isotropic case, at the onset of shear jamming~\cite{Sastry-PRE,Jin2021PNAS}; however, clear deviations are found in steady-states generated by cyclic shear~\cite{Wang2022PNAS}. 
A natural proposition is that, the values of these exponents, as well as their relationship that reflects the mechanical stability of the system, could be changed by the plastic rearrangements of the configuration during shear. 
However, this proposition still needs to be systematically examined.

% Summary
Here, by studying two-dimensional (2D) bidisperse jammed photoelastic disks subject to  quasi-static pure shear, we reveal the nature of yielding as a state point simultaneously displaying marginal stability and criticality. 
Analyzing the variation of the relationship between $\theta$ and $\gamma$ with the strain $\epsilon$, we find that, remarkably, during shear only two states are  marginally stable: the isotropically jammed state at $\epsilon=0$ and  the state at the yielding transition $\epsilon_{Y}$. 
Beyond $\epsilon_Y$, the system, which is in the steady state, is globally unstable.
An additional transition at $\epsilon_P$ can be identified before yielding, representing the onset of microstructure anisotropy and the location of maximum stability. 
The spatial correlations and global fluctuations of glass overlap order parameters suggest that yielding is critical where the corresponding susceptibilities peak.

%%%%%%%%%%%%%%%%%%%% Results
\section*{II. Results}
The average global stress ratio $\left \langle \tau/p \right \rangle$ (as defined in Methods) versus strain $\epsilon$ is shown in Fig.~\ref{fig:figure1}(c) for systems of different packing fractions $\phi's$, where the data points of different $\phi's$ collapse nicely onto a single master curve. Therefore, the behavior of the system's mechanical anisotropy depends weakly on the packing fraction $\phi$. For small strains, $\left \langle \tau/p \right \rangle$ increases approximately linearly with $\epsilon$. When $\epsilon>0.005$, $\left \langle \tau/p \right \rangle$ deviates from the linear increase with a gradual decreasing slope before reaching the maximum at $\epsilon\approx 0.03$, beyond which $\left \langle \tau/p \right \rangle$ starts decreasing and gradually levels off showing large fluctuations from data points of different $\phi's$.  
Some qualitative change occurs as $\epsilon$ increases, but clear transition points are obscured due to the ensemble-averaged smooth curve. 
A crucial question is whether there is a yielding transition with evident structural characteristics. In the following, we will explore the relationship between the power-law exponents associated with the weak-force and small-gap distributions, which are related to the mechanical stability, and the anisotropy and yielding of the system.

% exponents versus strain
\paragraph*{The exponents versus strain.} 
 
The measured exponents of the initial states are close to the predicted values, i.e., $\theta = 0.42311...$ and $\gamma = 0.41269...$ \cite{Charbonneau2014NC,Charbonneau2014JSMTE,Charbonneau2017ARCMP}, as shown in Fig.~\ref{fig:figure2}, which are also consistent with the early experiments\cite{Wang2022PNAS}. Meanwhile, the weak-force and small-gap distributions still exhibit power-law scalings while subject to shear, as shown in Fig.~\ref{fig:figure2}(a-b) for $\phi = 0.855$ and at different strains $\epsilon$ and the corresponding $\tau/p$. 
In Fig.~\ref{fig:figure2}, $f$ denotes the contact-force magnitude, and the dimensionless small interparticle gap is defined as $h = r_{ij}/ \left ( a_i+a_j \right )-1$, where $r_{ij}$ is the distance between particles $i$ and $j$, and $a_i$ and $a_j$ are the radii of the two particles, respectively. 
The insets of Fig.~\ref{fig:figure2}(a-b) show the exponents of $\theta$ and $\gamma$ versus $\epsilon$ obtained from the power-law fittings.  Distributions at other $\phi$ can be seen in SFig.1 of the Supplemental Materials.

We plot the measured exponent $\theta$ versus strain $\epsilon$ in Fig.~\ref{fig:figure2}(c) for various $\phi$, where data points of different $\phi's$ collapse nicely onto the same curve. We draw a smooth curve as a guide to the eye in the figure, showing the nonmonotonic change of $\theta$ versus $\epsilon$. Starting at $\epsilon=0$, the exponent $\theta$ decreases rapidly from the initial value of $\theta_0 \approx 0.42$ to the minimum $\theta_{\mathrm{min}}$, and then it slowly increases. Note that the variation of $\theta$ with $\epsilon$ shows an opposite trend to that of the stress ratio $\tau/p$ versus $\epsilon$.  We plot $\theta$ versus $\tau/p$ in the inset of Fig.~\ref{fig:figure2}(c), showing an approximately monotonic decreasing curve.

In Fig.~\ref{fig:figure2}(d), we plot $\gamma$ as a function of strain $\epsilon$ for different packing fractions with pressure $p>15\ \mathrm{N\ m^{-1}}$, where data points collapse nicely onto the same curve as drawn in a smooth curve for reference. The curve of $\gamma$ versus $\epsilon$ changes nonmonotonically with a peak centered around $\epsilon_P$, which, however, does not show apparent corresponding features in Fig.~\ref{fig:figure1}(c) on the curve of $\langle \tau/p\rangle$ or even on the curve of $\theta$ versus $\epsilon$ in Fig.~\ref{fig:figure2}(c).
To elucidate the physical meaning of this peak and to place the $\gamma$ and $\theta$ together for comprehensive comparison, we note that we can define $\gamma' = 1/\left ( 2+\theta \right )$ based on the scaling relationship between $\gamma$ and $\theta$ \cite{Charbonneau2014NC,Charbonneau2014JSMTE,Charbonneau2017ARCMP,Wyart2012PRL, Muller2015ARCMP}. 
We draw data points of $\gamma'$ versus $\epsilon$ and the associated smooth curve in Fig.~\ref{fig:figure2}(d). 
The direct comparison of $\gamma$ and $\gamma'$ shows several remarkable characteristics associated with the system's evolution under shear.
Firstly, before applying shear, $\gamma \approx \gamma'$ corresponds to the marginal stability of an isotropically jammed system, consistent with our early work \cite{Wang2022PNAS}.  
Secondly, the application of shear causes $\gamma$ to increase more rapidly with $\epsilon$ than $\gamma'$, reaching the peak at $\epsilon_P$ and then starting to decrease.  
Thirdly, the two exponents $\gamma$ and $\gamma'$ intersect again at $\epsilon_Y$, beyond which $\gamma$ still decreases at a faster rate, i.e., $\gamma < \gamma'$. 
Considering the fluctuations of data points, $\epsilon_P = 0.013 \pm 0.002$ and $\epsilon_Y = 0.030 \pm 0.005$ are defined from Fig.~\ref{fig:figure2}(d), which are drawn as the green and yellow vertical stripes in the corresponding figures.
The regime of $\gamma>\gamma'$ means that the system becomes stable under shear with the maximum stability obtained at $\epsilon=\epsilon_P$. In contrast, the $\gamma<\gamma'$ regime implies that the system becomes unstable under shear.
We identify the point $\epsilon_Y$ as the yielding point since it naturally separates the stable and unstable regimes, around which the $\langle \tau/p \rangle$ starts to decrease
according to Fig.~\ref{fig:figure1}(c). At $\epsilon_Y$, the system is marginally mechanically stable \cite{Wyart2012PRL, Muller2015ARCMP}. Only two points exhibit marginal stability along the shear curve: the unstrained isotropic jamming point and the yielding point, which is anisotropic and marks the emergence of global instability.

\paragraph*{The first point $\epsilon_P$: the occurrence of structural anisotropy.} 
Since shear introduces anisotropy, the isotropic analysis alone cannot fully reflect the system's properties. Thus, we investigate the probability distribution functions of the components of the weak contact forces and small gaps along the compression ($x$) and expansion ($y$) directions of pure shear. We define the gap vector $\mathbf{h} = \mathbf{r}_{ij}/\left ( a_i + a_j \right ) -  \mathbf{r}_{ij}/ \left | \mathbf{r}_{ij} \right |$, where $\mathbf{r}_{ij}$ is the relative position vector from particle $i$ to particle $j$, and the associated $x$ component $h_x$ and the $y$ component $h_y$. 
Since smaller gaps are more closely related to mechanical stability, only gaps of $h < 0.3$ are included in the statistics. 
Interestingly, the distributions of $h_x$ and $h_y$ also exhibit power laws, satisfying $g(h_x) \sim h_x^{-\gamma_x}$ and $g(h_y) \sim h_y^{-\gamma_y}$, as shown in SFig.2 of the Supplementary Materials.
Fig.~\ref{fig:figure3}(a) shows the exponents $\gamma_x$ and $\gamma_y$ versus strain $\epsilon$, where the data points of $\gamma$ are also plotted for comparison. 
The figure shows that micro structures develop clear anisotropy from $\epsilon_P$. Before $\epsilon_P$, $\gamma_x$ and $\gamma_y$ are indistinguishable and increase simultaneously with $\epsilon$, while after that, $\gamma_x$ increases further, but $\gamma_y$ starts to decline, and their difference becomes progressively large and eventually stabilizes after $\epsilon_Y$. Note that $\gamma \approx \gamma_x + \gamma_y - 1$ holds for all $\epsilon$, implying that the distributions of $h_x$ and $h_y$ are approximately independent.

The distributions of the contact-force components similarly exhibit power laws in the weak force regimes, satisfying $P(f_x) \sim f_x^{\theta_x}$ and $P(f_y) \sim f_y^{\theta_y}$, as shown in SFig.2 of the Supplementary Materials. However, unlike the gap, the exponents show anisotropy right at the beginning of the shear: $\theta_y$ decreases faster with strain than $\theta_x$, and after $\epsilon_P$ the differences between the two exponents are almost constant, as shown in Fig.~\ref{fig:figure3}(b) albeit with some small fluctuations among data points of different $\phi's$.

\paragraph*{The second point $\epsilon_Y$: the yielding transition.}
To reveal the criticality of the yielding point, we define an overlap order parameter $Q_{ab}$ between two independent samples (equation (1) in Methods), which is a similar approach compared to the replica overlap function $Q_{ab}$ introduced in ref.~\cite{Parisi2017PNAS,Procaccia2017PRE}. Three correlators $\Gamma_2$, $G_R$ and $G_L$ can also be defined by analogy to ref.~\cite{Parisi2017PNAS,Procaccia2017PRE} (See equations (2-5) in Methods).
Taking the system at $\phi = 0.847$ and $\epsilon = 0.040$ as an example, the 2D correlation functions $\Gamma_2(\mathbf{r})$ and $G_R(\mathbf{r})$ are shown in Fig.~\ref{fig:figure4}(a) and (b), respectively. $\Gamma_2$ decreases outward from the peak at the origin, and the correlations are stronger in the diagonal directions, which coincide with the directions of shear bands. The correlator $G_R$ is nonmonotonic along the diagonal direction, with a peak at each of the four corners.
 $G_L$ behaves like $G_R$. Additional information on these correlators is given in SFigs.(3-4) of the Supplemental Materials. 
 
 We set one of the diagonal directions as the $x'$ direction and the other perpendicular one as the $y'$ direction, as shown in Fig.~\ref{fig:figure4}(a) and (b). Since the correlators along the diagonal directions are the most sensitive, we define the susceptibilities $\chi_{\Gamma_2}^{x'}$, $\chi_{G_R}^{x'}$ and $\chi_{G_L}^{x'}$ as the integral of the correlators along the $x'$ direction (equation (6) in Methods). 
 %For example,
%\begin{equation}
%    \chi^{x'}_{\Gamma_2}(\epsilon) \equiv \int dx'\ \Gamma_2(x',y'=0;\epsilon).
%\end{equation}
These susceptibilities essentially quantify the global fluctuations of the overlap order parameter at different strains. 
In Fig.~\ref{fig:figure4}(c), we show the susceptibilities as functions of $\epsilon$ for systems of different $\phi's$. It is clear that there is a distinct peak in $\chi_{G_R}^{x'}$ and $\chi_{G_L}^{x'}$ at $\epsilon_Y$, which represents the criticality of the yielding transition.   $\chi_{\Gamma_2}^{x'}$ changes more gently, and its maximum is located at a larger strain. We also find that the strain where these susceptibilities begin to increase from a value close to zero is around $\epsilon_P$.

%%%%%%%%%%%%%%%%%%%% Discussion
\section*{III. Discussion}
% Conclusion
We have experimentally measured the two exponents $\theta$ and $\gamma$ related to mechanical stability as functions of strain $\epsilon$ in dense bidisperse photoelastic disk packings subject to pure shear. We have found that both $\theta$ and $\gamma$ evolve nonmonotonically with $\epsilon$, which allows us to identify two characteristic points of strain, i.e., $\epsilon_P$ and $\epsilon_Y$.  $\epsilon_P$ refers to the strain corresponding to the peak value of $\gamma$, and $\epsilon_Y$ refers to the strain of the intersection point between $\gamma$ and $\gamma'=1/(2+\theta)$. The first point $\epsilon_P$ is associated with the emergence of anisotropy of microstructures. The second point $\epsilon_Y$ characterizes the yielding transition, which is marginally mechanically stable \cite{Wyart2012PRL,Muller2015ARCMP, lerner2013low, degiuli2014force} 
 with its criticality characterized by the overlap order parameter and correlators defined analogously to the replica methods. These findings demonstrate that the plasticity and yielding are closely associated with the qualitative microstructural changes as characterized by $\theta$ and $\gamma$: starting from a marginally stable initial state, the system first becomes stable at small strains and then eventually becomes unstable as shear continues; the transition point is the yield, where the system satisfies the scaling relationship between $\theta$ and $\gamma$, i.e., $\gamma=\gamma'$, despite with strong anisotropy in their corresponding $x$ and $y$ components and with the system-spanning shear band (See, e.g., SFig.5 in the Supplemental Materials).

This enhancement in stability due to shearing is reminiscent of shear-induced dilatancy: the system is stabilized with a fixed volume or dilates with a fixed pressure.
Meanwhile, the number of plastic events increases with strain, and the plasticity becomes dominant at $\epsilon_P$, where the stability starts decreasing. When the number of plastic events continues increasing, these plastic events in the form of Eshelby quadrupoles concatenate each other, forming distinct shear bands at $\epsilon_Y$. See SFig.5 of the Supplemental Materials for the specific spatial distribution of the plastic events.

In this paper, we find that there is a characteristic point $\epsilon_P$ before yielding. We note that in Refs.\cite{Rainone2015PRL, Jin2018SA} it has  been pointed out that there is a Gardner transition before yielding in a hard-sphere glass system, which separates stable (reversible) and marginally stable (partially irreversible) glass phases. 
Since $\epsilon_P$ is in the stable phase, it cannot be explained by the picture of a Gardner transition. 
Moreover, in granular materials, friction makes the deformation always irreversible: even if the configuration may be reversible, the corresponding force network cannot. 
The physical origin of $\epsilon_P$ thus remains to be theoretically understood.

Note that despite the observed criticality and the system-spanning shear band near the yielding, it is still an open question whether the yielding is a first-order thermodynamic phase transition \cite{Jaiswal2016PRL, Kawasaki2016PRE}, a second-order  thermodynamic phase transition\cite{Ozawa2018PNAS}, or a spinodal point \cite{Rainone2015PRL, Parisi2017PNAS, Procaccia2017PRE}. 
Granular matter differs from molecular or colloidal glasses due to 
its athermal and frictional characteristics.
The remaining challenge is to perform  finite-size and scaling analyses in the vicinity of yielding, which would require a large number of samples and sufficient experimental accuracy. 

%%%%%%%%%%%%%%%%%%%% Experimental Setup
\section*{Methods}
\subsection{Experimental apparatus}
The 2D granular system is composed of $2,710$ small disks and $1,355$ large disks, whose diameters are $d_s = 10\ \mathrm{mm} = \mathrm{d}$ and $d_l = 14\ \mathrm{mm} = 1.4\ \mathrm{d}$, respectively. 
Disks are placed within a rectangular area enclosed by two pairs of walls on top of a glass plate. 
Each pair can move freely to apply isotropic compression or area-conserved pure shear. 
Eight mini vibrators are attached to the edge of the glass plate and synchronized with wall movement, providing vibrations to eliminate the base friction. 
At the top is a high-resolution ($10\ \mathrm{pixels/mm}$) $2\times2$ array of cameras for image acquisition. One circular polarizer below the cameras can move in and out of the field of view as needed, and the other matched polarizer is attached under the glass plate, with a green LED light source below. More details can be found in ref. \cite{Wang2020NC,Wang2021PRR,Wang2022PNAS}.

\subsection{Initial state preparation}
An initial state of pure shear of a given $\phi$ is prepared as follows. Firstly, we prepare a stress-free random and homogeneous configuration at $\phi = 0.834$ below the jamming point $\phi_J \approx 0.84$ of frictionless particles by compressing a loose random configuration while gently perturbing particles to destroy any transient force chains and meanwhile using mini vibrators to eliminate the base friction. Secondly, the stress-free system is subjected to quasi-static isotropic compression while using mini vibrators to eliminate the base friction till reaching the target packing fraction $\phi$. The above protocol yields an isotropic and homogeneous initial state of force chains, as shown, e.g., in Fig.~\ref{fig:figure1}(a) with $\phi = 0.855$.  Applying shear causes the system's pressure $p$ to increase with strain due to shear dilatancy. 
Since the contact force resolution is around $0.05\ \mathrm{N}$, a precise power law interval of the weak force distribution can be best resolved when $p>15\ \mathrm{N\ m^{-1}}$. Moreover, the accuracy of force measurement decreases when $p>30\ \mathrm{N\ m^{-1}}$. 
Therefore, to achieve the best accuracy, it requires $p \in [15\ \mathrm{N\ m^{-1}}, 30\ \mathrm{N\ m^{-1}}]$. 
To obtain the weak-force critical exponent $\theta$ for a wide range of strain $\epsilon$, we select seven different $\phi's$, in between $\phi_{\mathrm{min}} = 0.840$ and $\phi_{\mathrm{max}} = 0.859$. 
The inset of Fig.~\ref{fig:figure1}(c) shows $p$ versus $\epsilon$ for various $\phi's$, where distinct power law distributions of weak forces can be accurately measured when the pressure is above the dashed line.

\subsection{Pure shear protocols}
After preparing the initial state at the given $\phi$, we apply area-conserved quasi-static pure shear to the system by compressing along the $x$ direction while expanding along the $y$ direction in a series of incremental strain steps. The maximum strain is set differently to adapt to the corresponding $\phi$ to ensure pressure $p$ is below $30\ \mathrm{N\ m^{-1}}$ for the accurate measurement of contact forces. The shear strain $\epsilon \equiv |\mathrm{d}x|/x_0$, where $x_0$ is the initial system size along the compression direction and $|\mathrm{d}x|$ is the corresponding change in linear size. Fig.~\ref{fig:figure1}(b) shows the contact force network at $\epsilon = 0.016$ after applying pure shear to the initial state in Fig.~\ref{fig:figure1}(a). For each $\phi$, we perform 20 independent experimental runs.

\subsection{Stress measurement}
At each strain $\epsilon$, we record one stress image (as shown in Fig.~\ref{fig:figure1}(a-b)) and one normal image without the polarizer in front of the cameras to detect the disk positions. The contact forces can be measured with an accuracy of $5\%$ using a force-inverse algorithm detailed in \cite{Wang2021PRR, Wang2022PNAS}. From contact forces, the system's stress tensor $\hat{\sigma}$ can be defined as 
 $ \hat{\sigma} = \frac{1}{S} \sum_{i \ne j} \mathbf{r}_{ij} \otimes \mathbf{f}_{ij}$ ,
where $S$ is the system's area, $\mathbf{r}_{ij}$ is the contact vector from the center of the particle $i$ to the contact point between particle $i$ and $j$, $\mathbf{f}_{ij}$ is the force vector associated with this contact and $\otimes$ denotes the vector outer product. 
We denote the principal stresses of $\hat{\sigma}$ by $\sigma_1$ and $\sigma_2$, and the pressure $p = \left ( \sigma_1 + \sigma_2 \right ) /2$ and the shear stress $\tau = \left | \sigma_1 - \sigma_2 \right | /2$. The angle brackets $\left \langle \cdot \right \rangle$ indicate an ensemble average over the 20 experimental data sets. To avoid the boundary effect, we remove the particles within $10d$ of the boundaries in the analysis.

\subsection{Correlation functions and susceptibilities}
As an analogy to ref.\cite{Parisi2017PNAS,Procaccia2017PRE}, we first consider the initial states of two independent samples $a$ and $b$, numbering all particles in $a$ and denoting the position of the $i$th particle as $\mathbf{r}_i^a$. Next, we number all particles in sample $b$ in the sorted order such that the particle closest to the $i$th particle in $a$ is marked as the $i$th particle in $b$. 
After numbering the particles in the initial states, we track them when the strain is applied while keeping their labeling numbers fixed. An analogous "overlap" function is defined as
\begin{equation}
  Q_{ab}(\epsilon) = \frac{1}{N} \sum_{i=1}^{N} \theta ( l - |(\mathbf{r}_i^a(\epsilon)-\mathbf{r}_i^a(0)) - (\mathbf{r}_i^b(\epsilon)-\mathbf{r}_i^b(0))| ),
\end{equation}
where $N$ is the total number of the particles, $\theta(x)$ is the Heaviside step function, and $l$ is a constant length chosen as $l = 0.3\mathrm{d}$ according to previous work \cite{Berthier2013PRE,Jaiswal2016PRL,Parisi2017PNAS,Procaccia2017PRE}. This definition of "overlap" describes the similarity of the local deformation caused by shearing between independent isotropically jammed packings.

The correlators can also be defined by analogy according to ref. \cite{Parisi2017PNAS,Procaccia2017PRE}, as
\begin{equation}
\begin{split}
 & \Gamma_2 (\mathbf{r}) \equiv \\
 & \left \langle \frac{\sum_{i \neq j} (u_i^{ab}-Q_{ab})(u_j^{ab}-Q_{ab})\delta(\mathbf{r}-(\mathbf{r}_i^a - \mathbf{r}_j^a))}{\sum_{i \neq j} \delta(\mathbf{r}-(\mathbf{r}_i^a - \mathbf{r}_j^a))} \right \rangle, 
\end{split}
\end{equation}
\begin{equation}
\begin{split}
& G_R (\mathbf{r}) \equiv \\
& \left \langle \frac{\sum_{i \neq j} (u_i^{ab}u_j^{ab} - 2u_i^{ab}u_j^{ac} + Q_{ab}Q_{cd})\delta(\mathbf{r}-(\mathbf{r}_i^a - \mathbf{r}_j^a))}{\sum_{i \neq j} \delta(\mathbf{r}-(\mathbf{r}_i^a - \mathbf{r}_j^a))} \right \rangle, 
\end{split}
\end{equation}
and
\begin{equation}
    G_L (\mathbf{r}) \equiv 2G_R (\mathbf{r}) - \Gamma_2 (\mathbf{r}), 
\end{equation}
with
\begin{equation}
    u_i^{ab} \equiv \theta ( l - |(\mathbf{r}_i^a(\epsilon)-\mathbf{r}_i^a(0)) - (\mathbf{r}_i^b(\epsilon)-\mathbf{r}_i^b(0))| ).
\end{equation}
Here, we use $\left \langle \cdot \right \rangle$ to denote the average of all possible combinations of total samples. And we define the corresponding susceptibilities as the integral of the correlators along one of the diagonal directions ($x'$ direction). For example,
\begin{equation}
    \chi^{x'}_{\Gamma_2}(\epsilon) \equiv \int dx'\ \Gamma_2(x',y'=0;\epsilon).
\end{equation}

%%%%%%%%%%%%%%%%%%%%%%%%%%%%%%%%%%%%%%%%%%%%%%%%%%%%%%%%%%%%%%%%%%%%%%%%%%%%%%%%%%%%%%%
\begin{acknowledgments}
This work is supported by the NSFC (No. 11974238 and No. 12274291). This work is also supported by the Innovation Program of Shanghai Municipal Education Commission under No. 2021-01-07-00-02-E00138. We also acknowledge the support from the Student Innovation Center of Shanghai Jiao Tong University. Y.Q.W. acknowledges support from Shanghai Jiao Tong University via the scholarship for outstanding Ph.D. graduates. 
Y.J. acknowledges support from NSFC (Grants
11974361, 12161141007, 11935002, and 12047503), from 
Chinese Academy of Sciences (Grants ZDBS-LY-7017 and KGFZD-145-22-13), 
and from Wenzhou Institute (Grant WIUCASICTP2022).

\end{acknowledgments}

\bibliography{references}

\end{document}

\begin{CJK*}{UTF8}{gbsn}
\title{The yielding of granular matter is marginally stable and critical:\\ Supplementary Materials}

\author{Jin Shang}
    \affiliation{School of Physics and Astronomy, Shanghai Jiao Tong University, 800 Dong Chuan Road, 200240 Shanghai, China.}
\author{Yinqiao Wang}
    \affiliation{Research Center for Advanced Science and Technology, University of Tokyo, 4-6-1 Komaba, Meguro-ku, Tokyo 153-8505, Japan.}
\author{Yuliang Jin}
    \affiliation{Institute of Theoretical Physics, Chinese Academy of Sciences, Beijing 100190, China. }
    \affiliation{School of Physical Sciences, University of Chinese Academy of Sciences, Beijing 100049, China.}
    \affiliation{Wenzhou Institute, University of Chinese Academy of Sciences, Wenzhou 325000, China.}
\author{Jie Zhang}
    \email[Email address: ]{jiezhang2012@sjtu.edu.cn}
    \affiliation{School of Physics and Astronomy, Shanghai Jiao Tong University, 800 Dong Chuan Road, 200240 Shanghai, China.}
    \affiliation{Institute of Natural Sciences, Shanghai Jiao Tong University, 200240 Shanghai, China.}
    
\maketitle
\end{CJK*} 

%%%%%%%%%%%%%%%%%%%%%%%%%%%%%%%%%%%%%%%%%%%%%%%%%
\begin{figure*}
	\centerline{\includegraphics[width = 18 cm]{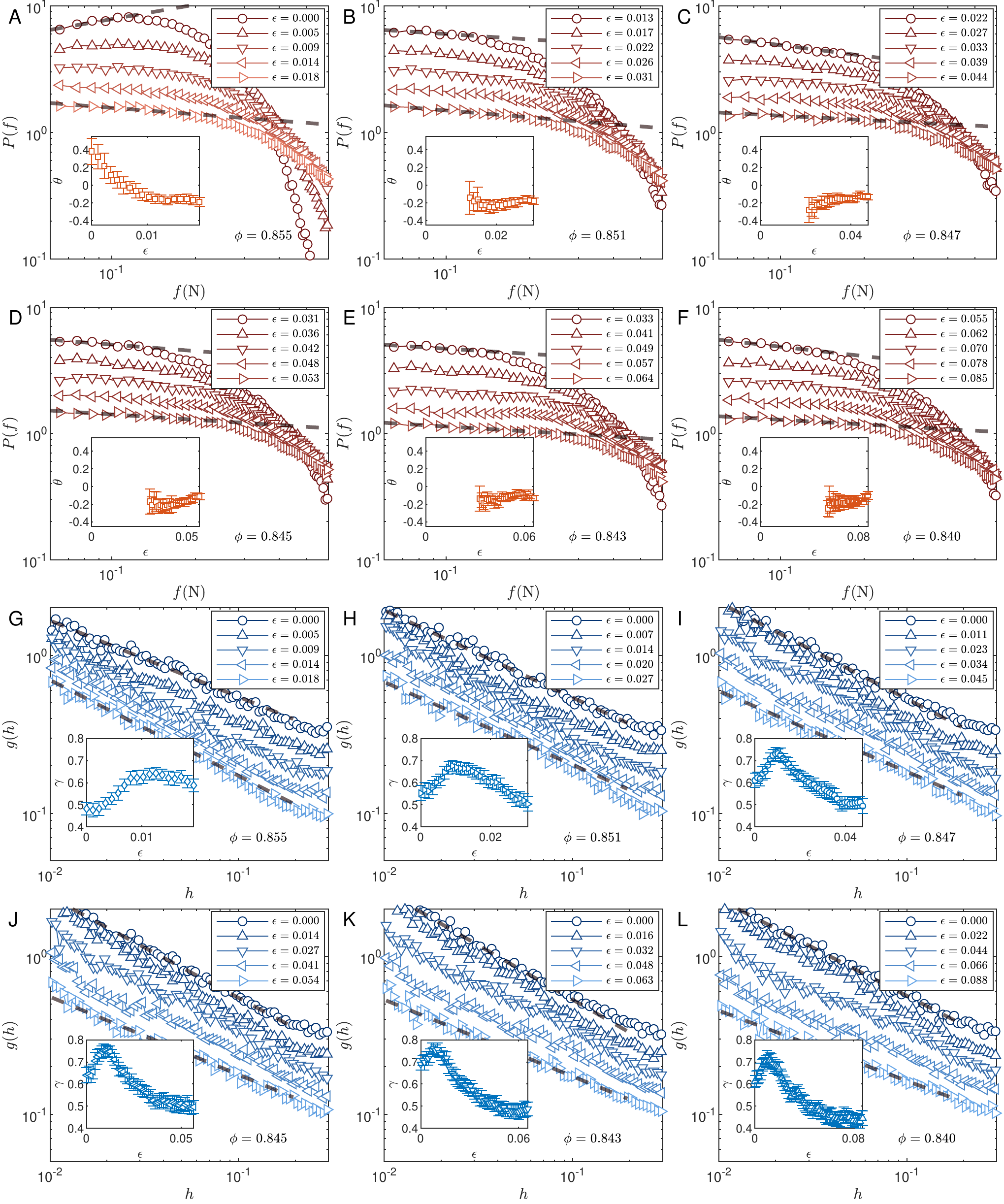}}
	\caption{
            (a-f) Contact-force distributions $P(f)$ of different packing fractions $\phi's$. Insets: The weak-force critical exponents $\theta$ versus the shear strain $\epsilon$. (g-l) Small-gap distributions $g(h)$ of different $\phi's$. Insets: The critical small-gap exponents $\gamma$ versus $\epsilon$. Two dashed lines in each panel represent the power-law fitting curves associated with the maximum and minimum strains, respectively. The error bars in the insets represent the standard deviations of the coefficients of the fitting parameters.
            }
	\label{fig:sfigure1}
\end{figure*}

\begin{figure*}
	\centerline{\includegraphics[width = 18 cm]{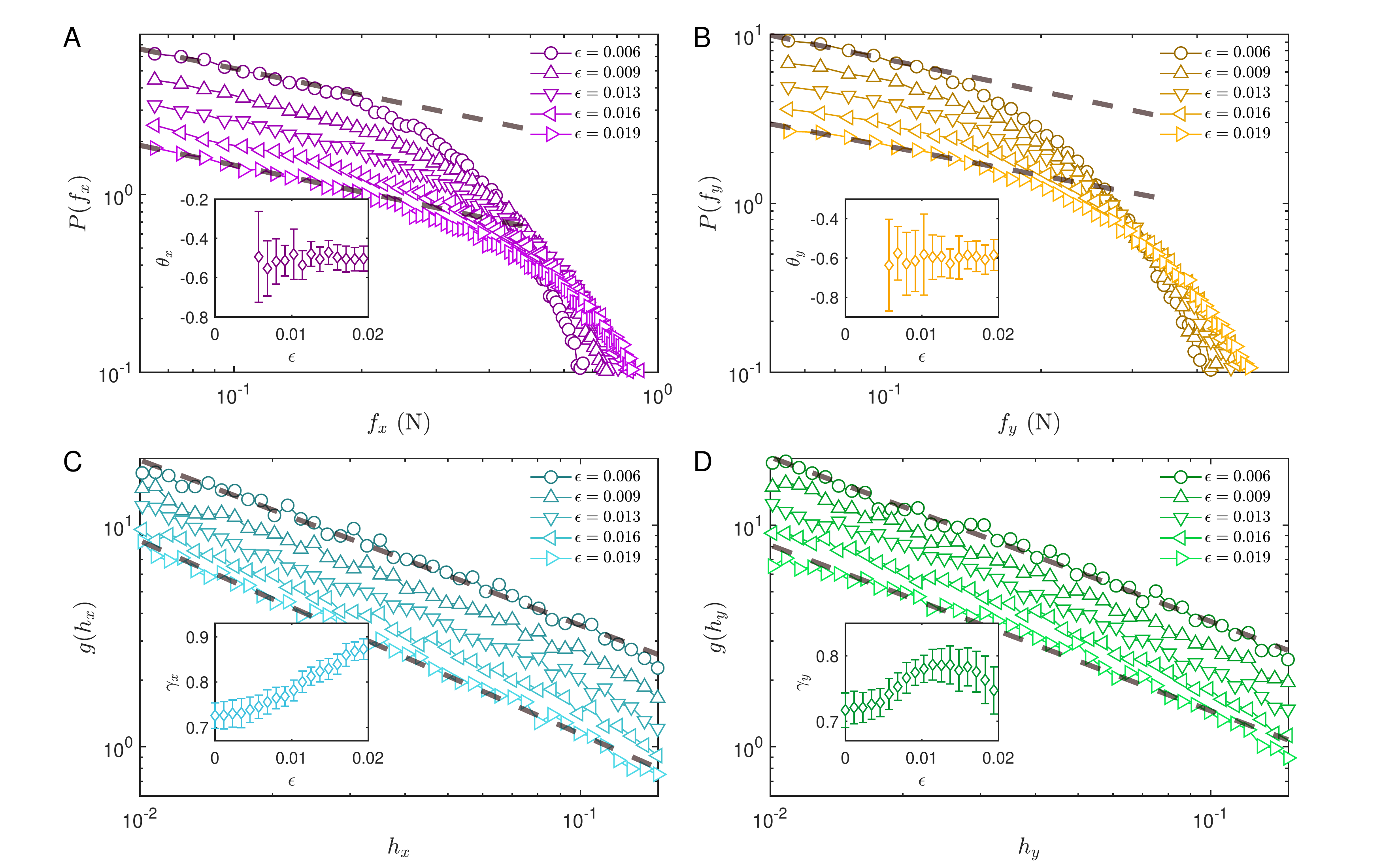}}
	\caption{
            Distributions of the $x$ and $y$ components of contact forces $f_x$ (a), $f_y$ (b) and small gaps $h_x$ (c), $h_y$ (d) of $\phi = 0.855$. Insets: the corresponding critical exponents $\theta_x$, $\theta_y$, $\gamma_x$ and $\gamma_y$ versus $\epsilon$. Two dashed lines in each panel represent the power-law fitting curves associated with the maximum and minimum strains, respectively. The error bars in the insets represent the standard deviations of the coefficients of the fitting parameters.
            }
	\label{fig:sfigure2}
\end{figure*}

\begin{figure*}
	\centerline{\includegraphics[width = 18 cm]{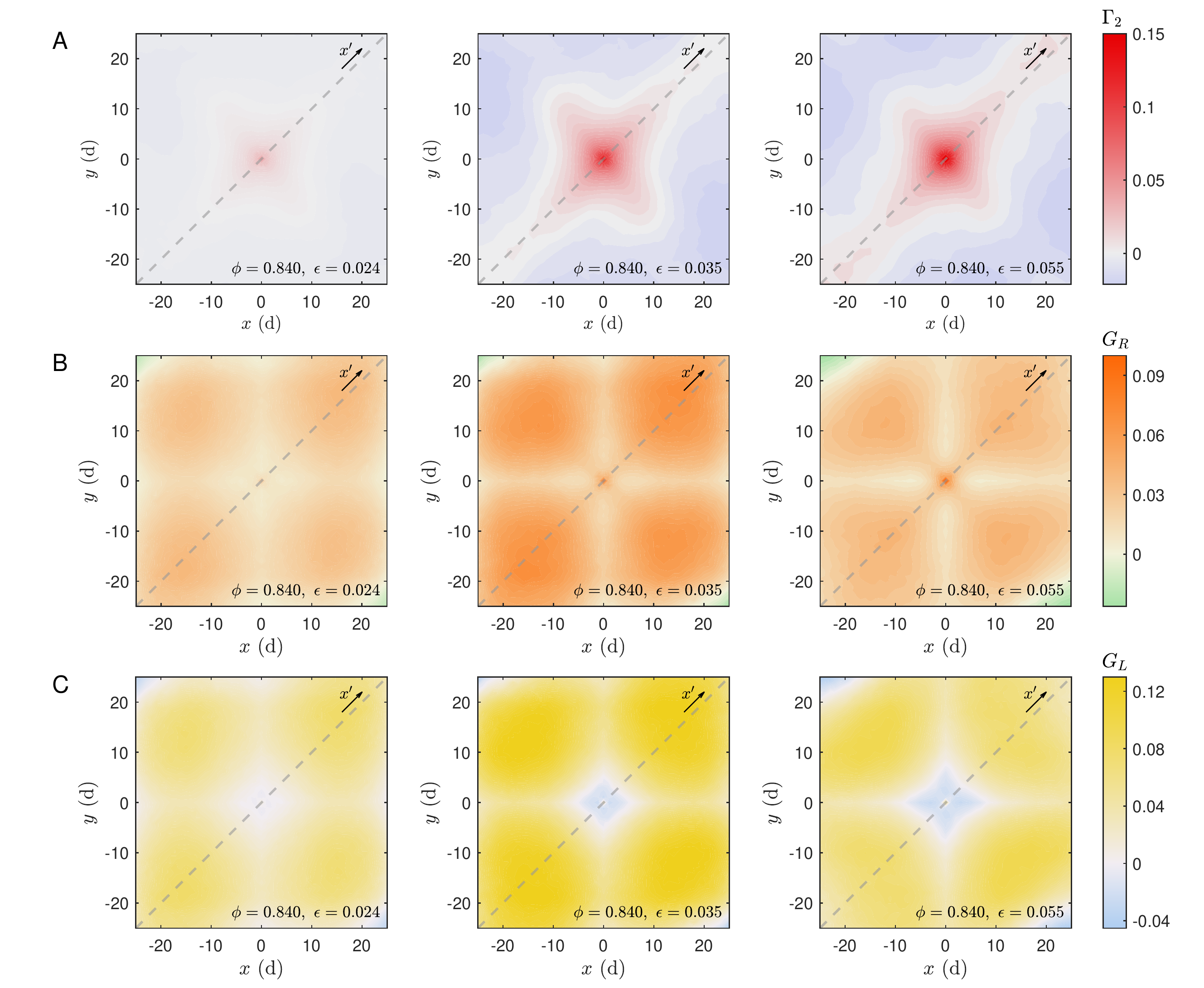}}
	\caption{
            Examples of the 2D correlation functions $\Gamma_2 (x,y)$ (a), $G_R (x,y)$ (b) and $G_L (x,y)$ (c). Each column contains the cases before, close to, and after yielding. The arrows specify the $x'$ direction, and the dashed lines represent $y'=0$.
            }
	\label{fig:sfigure3}
\end{figure*}

\begin{figure*}
	\centerline{\includegraphics[width = 18 cm]{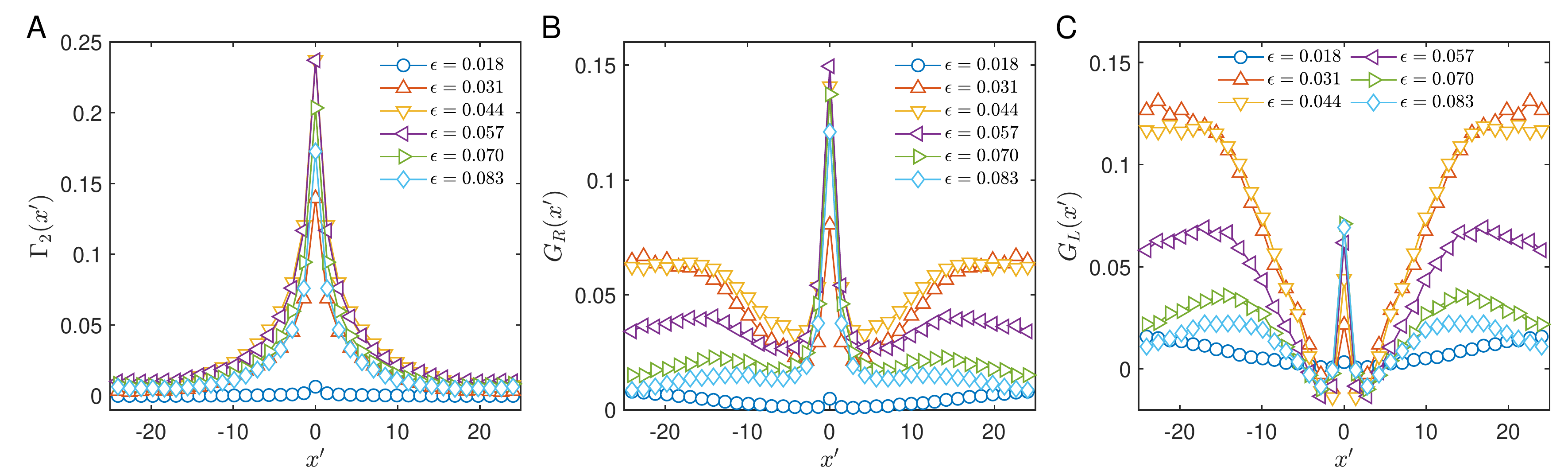}}
	\caption{
            Cuts of correlation functions $\Gamma_2 (x')$ (a), $G_R (x')$ (b) and $G_L (x')$ (c) along the line $y'=0$ at different shear strains $\epsilon's$.
            }
	\label{fig:sfigure4}
\end{figure*}

 \begin{figure*}
	\centerline{\includegraphics[width = 18 cm]{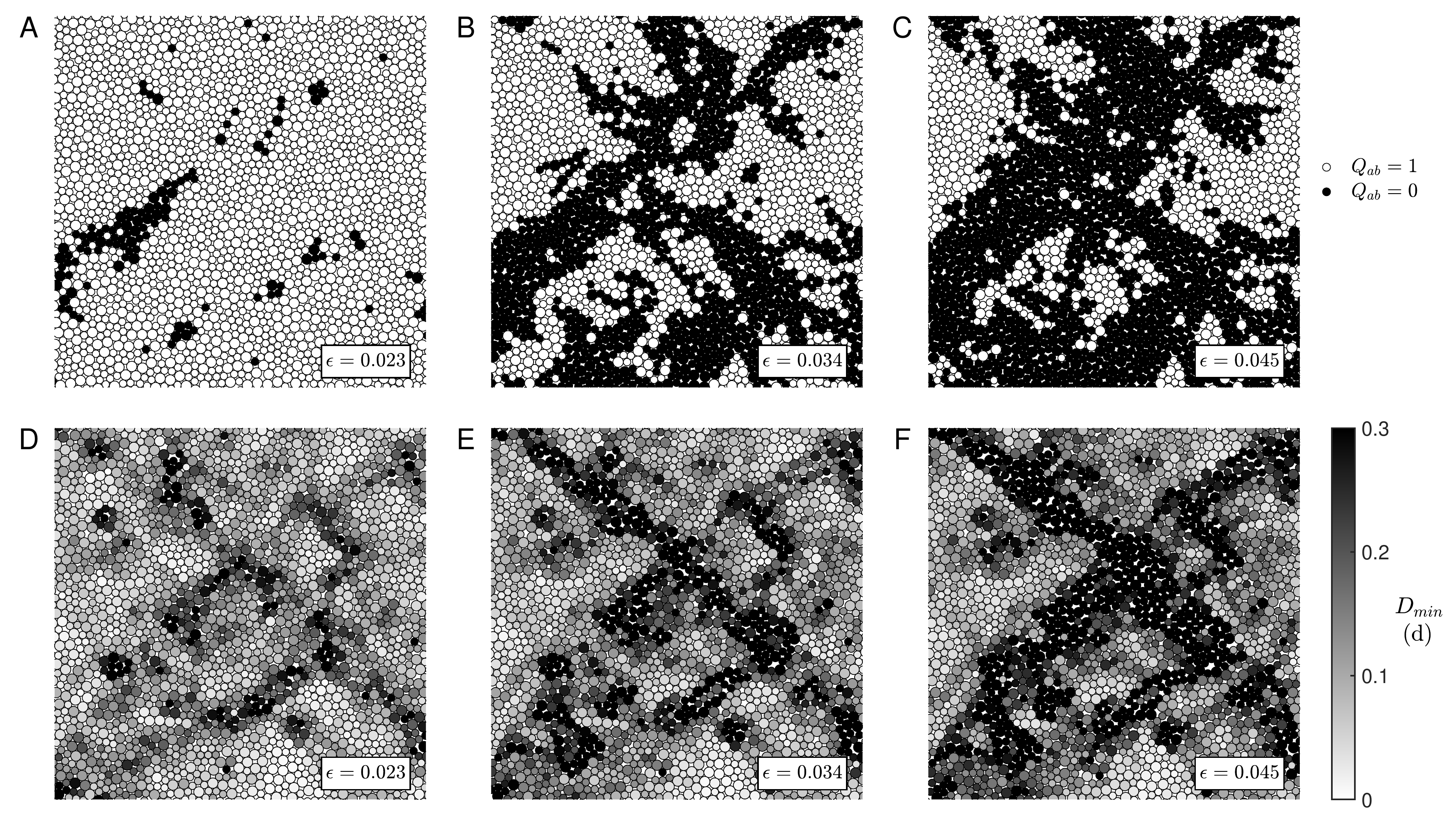}}
	\caption{
            Examples of spatial distributions of the local overlap order parameter $Q_{ab}$ (a-c) and the local non-affinity $D_{min}$ (d-f) at $\phi=0.845$. 
            }
	\label{fig:sfigure5}
\end{figure*}

%%%%%%%%%%%%%%%%%%%%%%%%%%%%%%%%%%%%%%%%%%%%%%%%%%
In the Supplemental Materials, we describe in more detail the distributions of contact forces and small gaps of different packing fractions $\phi's$, the distributions of the $x$ and $y$ components of contact forces and small gaps of $\phi = 0.855$, the 2D correlation functions of $\Gamma_2$, $G_R$, and $G_L$ as defined in the main text and the related cuts along the $ x'$ direction, and the spatial distributions of plastic events in terms of local overlap order parameter $Q_{ab}$ and local nonaffinity $D_{min}$. 

SFig.~\ref{fig:sfigure1} shows the contact-force distributions and the small-gap distributions of various volume fractions, where the results of $\phi = 0.859$ are not included since they are already established in the main text. The corresponding critical exponents $\theta$ and $\gamma$ are shown in the insets. 
In SFig.~\ref{fig:sfigure1}, the force-distribution curves corresponding to pressure smaller than $15\ \mathrm{N\ m^{-1}}$ are not drawn to achieve the best measurement accuracy, as discussed in the main text. Examples of the distributions of the $x$ and $y$ components of contact forces and small gaps are shown in SFig.~\ref{fig:sfigure2}. The distributions of the $x$ and $y$ components of weak forces and small gaps exhibit power laws as $P(f_x) \sim f_{x}^{-\theta_x}$, $P(f_y) \sim f_y^{-\theta_y}$, $g(h_x) \sim h_x^{\gamma_x}$ and $g(h_y) \sim h_y^{\gamma_y}$. The insets show the corresponding power-law exponents versus shear strain $\epsilon$.

SFig.~\ref{fig:sfigure3} shows the two-dimensional (2D) correlation functions $\Gamma_2 (x,y)$, $G_R(x,y)$, and $G_L(x,y)$ in the cases before, close to, and after yielding. SFig.~\ref{fig:sfigure4} shows the cuts of the correlators along the diagonal line $y'=0$ at different $\epsilon's$, where their strength rises rapidly before yielding and then falls back afterward.

Examples of the specific spatial distributions of the plastic events are shown in SFig.~\ref{fig:sfigure5}. In SFig.~\ref{fig:sfigure5}(a-c), the plastic events are characterized by the localization of overlap order parameter defined in the main text, i.e.
\begin{equation}
    Q_{ab}(\epsilon; \mathbf{r}) \equiv \theta ( l - |(\mathbf{r}_i^a(\epsilon)-\mathbf{r}_i^a(0)) - (\mathbf{r}_i^b(\epsilon)-\mathbf{r}_i^b(0))| ) \delta(\mathbf{r}-\mathbf{r}_i^a(\epsilon)). 
\end{equation}
 Moreover, we use the local non-affinity $D^2_{min}$ of the particle displacements to characterize local plastic events as defined in ref.\cite{Falk1998PRE}. This $D^2_{min}$ represents the minimum mean-square difference between the actual displacement of the nearest neighbor particles relative to a center particle and their affine displacement if they would move with a uniform strain.\cite{Falk1998PRE}. In SFig.~\ref{fig:sfigure5}(d-f), we plot the spatial distributions of $D_{min} \equiv \sqrt{D^2_{min}}$ for the convenience of comparison. In both representations, it can be seen that the plastic events are much more broken and scattered before yielding. These plastic events are spatially percolated near yielding and form obvious shear bands, further strengthened and enlarged after yielding.

\bibliography{sreferences}